\documentstyle[lprocldpf,epsf]{article}

\def\pmb#1{\setbox0=\hbox{#1}%
  \kern-.025em\copy0\kern-\wd0
  \kern.05em\copy0\kern-\wd0
  \kern-.025em\raise.0433em\box0 }

\def\lb{\pmb{$\ell$}}

\def\txN{{\textstyle{1\over N}}}
\def\txN2{\textstyle{1\over N^2}}

\def\im{{\rm Im\,}}

\def\cuo2{CuO$_2$}

\def\2kf{2\,k_F}
\def\4kf{4\,k_F}

\def\q{{\bf q}}
\def\p{{\bf p}}

\def\om{\omega}

\def\Journal#1#2#3#4{{#1} {\bf #2}, #3 (#4)}

\def\PRL{\em Phys. Rev. Lett.}
\def\PRB{{\em Phys. Rev.} B}
 
\def\p{{\bf p}}
\def\q{{\bf q}}
\def\r{{\bf r}}

\def\om{\omega}
\def\xn{\langle n\rangle}

\def\gtwid{\mathrel{\raise.3ex\hbox{$>$\kern-.75em\lower1ex\hbox{$\sim$}}}}
\def\ltwid{\mathrel{\raise.3ex\hbox{$<$\kern-.75em\lower1ex\hbox{$\sim$}}}}

\begin{document}

\title{LOW-ENERGY ELECTRONIC EXCITATIONS OF THE LAYERED CUPRATES\\
AND THE HUBBARD MODEL}

\author{N. BULUT}

\address{Department of Physics, University of California,
Santa Barbara,\\ CA 93106--9530, USA}


\maketitle\abstracts{
We present a review of the Quantum Monte Carlo results 
on the magnetic, charge and single--particle excitations 
as well as the pairing correlations of the two--dimensional Hubbard model.
We are particularly interested in how these quantities are 
related to each other
in the metallic state that forms near half--filling.
These results are useful in gaining a better understanding of the 
low energy excitations of the superconducting cuprates.
}

\section{Introduction}

Superconducting cuprates have many anomalous physical 
properties \cite{Houston}.
The parent compounds are antiferromagnetic (AF) charge--transfer 
insulators.  
Upon finite doping long--range order is destroyed, and a
new strongly--correlated metallic state 
with unusual transport properties forms.
In this state there are strong short--range AF correlations.
Angular resolved photoemission experiments show that 
the single--particle properties are heavily damped and strongly 
renormalized. 
Charge dynamics probed by various transport measurements
are also unusual.
Most importantly,
a large number of experiments point out that the order parameter
of the superconducting state has $d_{x^2-y^2}$ 
symmetry \cite{DJS}.

Perhaps, the simplest model that has similar electronic
properties is the two--dimensional single--band Hubbard model
given by
\begin{equation}
H=-t\sum_{\langle ij\rangle ,s} \left( c^\dagger_{is}c_{js} +
c^\dagger_{js}c_{is} \right) + U \sum_i n_{i\uparrow}n_{i\downarrow}.
\label{eq:ham}
\end{equation}
Here $c^\dagger_{is}$ creates an electron of spin $s$ on site $i$, in
the first term the sum is over near-neighbor sites, and $n_{is} =
c^\dagger_{is}c_{is}$ is the occupation number for
electrons with spin $s$ on site $i$.
The near--neighbor hopping matrix element is $t$, and 
the onsite Coulomb repulsion is $U$.

Here, we will present Quantum Monte Carlo results 
on the magnetic, charge and single--particle excitations, 
and the pairing correlations of the Hubbard model
in the intermediate coupling regime near half--filling
for an $8\times 8$ lattice.
We will see that a strongly correlated metallic band develops upon  
doping the antiferromagnetic Mott--Hubbard insulator.
This metallic state has anomalous electronic properties.
For instance, the quasiparticle properties are 
strongly renormalized.
The system has strong short--range and low--frequency
AF fluctuations, and 
the long wavelength charge response is enhanced.
Furthermore, the system exhibits singlet $d_{x^2-y^2}$ 
pairing correlations \cite{SLH,Bickers}.
There have been detailed Quantum Monte Carlo and exact diagonalization 
studies of these properties of the strongly correlated Hubbard 
or $t$--$J$ models.
Here, we are especially interested in how these
properties are related to each other.
We will see that as the AF fluctuations grow, 
a narrow quasiparticle band, which has a large fermi surface,
develops at the top of the 
lower Hubbard band.
The AF fluctuations are also reflected in the effective 
particle--particle interaction, which is strongly repulsive
at large momentum transfers.
This feature of the effective interaction along with 
the large fermi surface causes $d_{x^2-y^2}$ 
pairing correlations \cite{SLH,Bickers,DJS2}.
We think that
these numerical results are useful in gaining a better understanding 
of the nature of the low--energy electronic excitations in the 
superconducting cuprates.

\section{Magnetic Fluctuations}

\begin{figure} 
\centerline{\epsfysize=6.8cm \epsffile[-30 184 544 598] {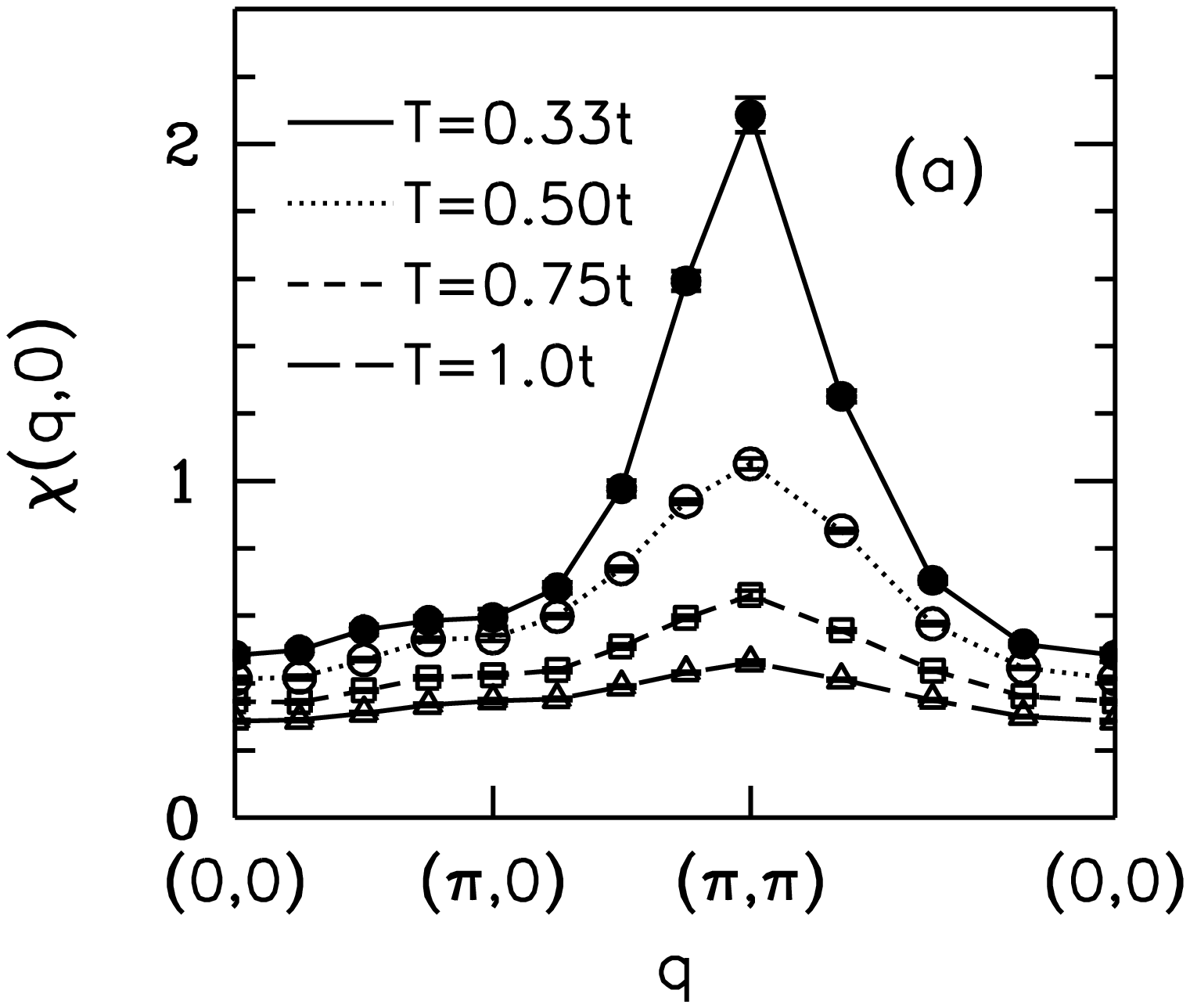}
\epsfysize=6.8cm \epsffile[98 184 672 598] {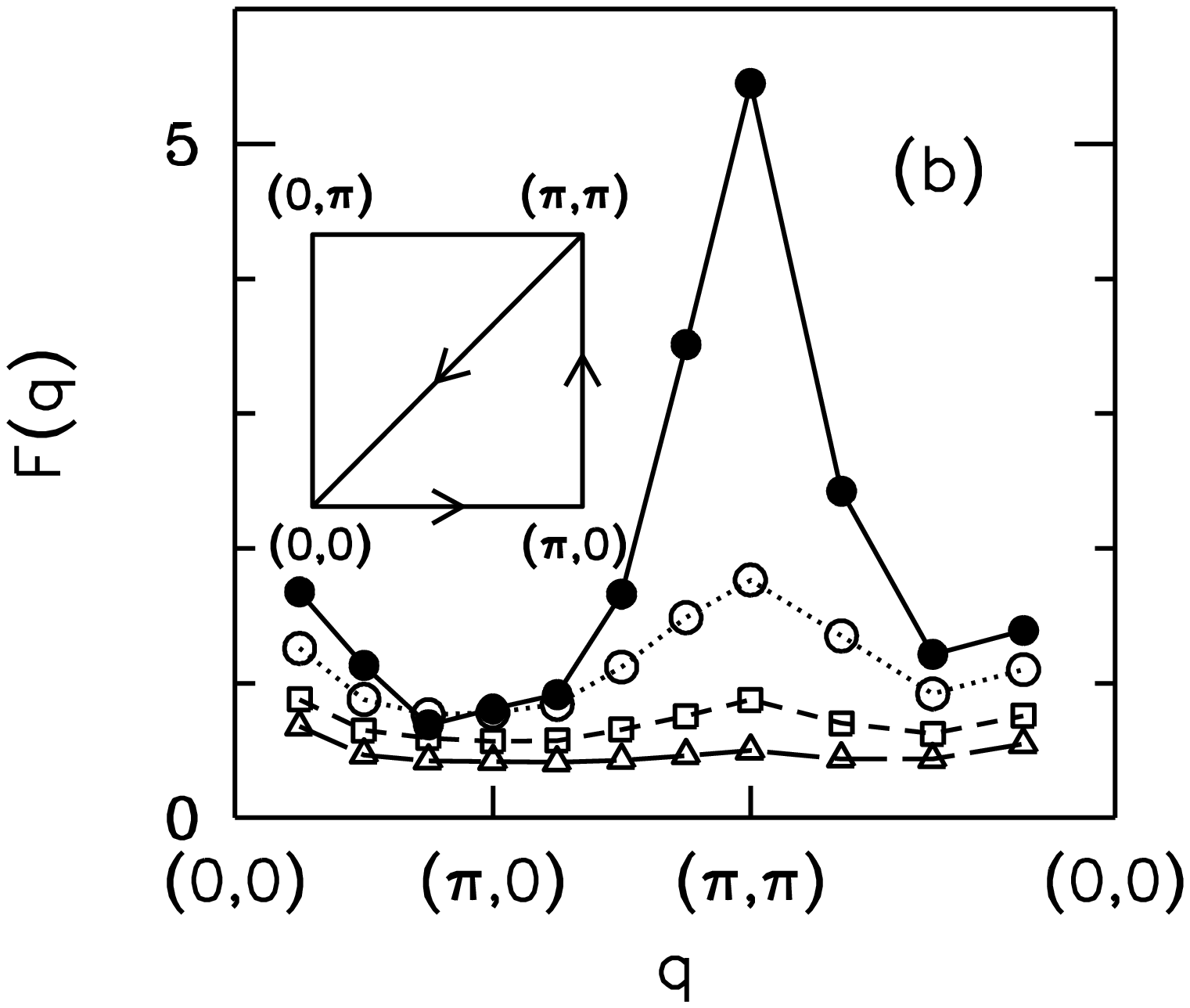}}
\vspace{0.3cm}
\caption{
Momentum dependence of (a) $\chi(\q,i\om_m=0)$ and 
(b) $F(\q)$ for $U/t=8$ and $\xn=0.875$.
Here $\q$ is plotted along the path show in the inset of (b).
\label{fig:chiq}}
\end{figure}

\begin{figure} 
\centerline{\epsfysize=6.8cm \epsffile[-30 184 544 598] {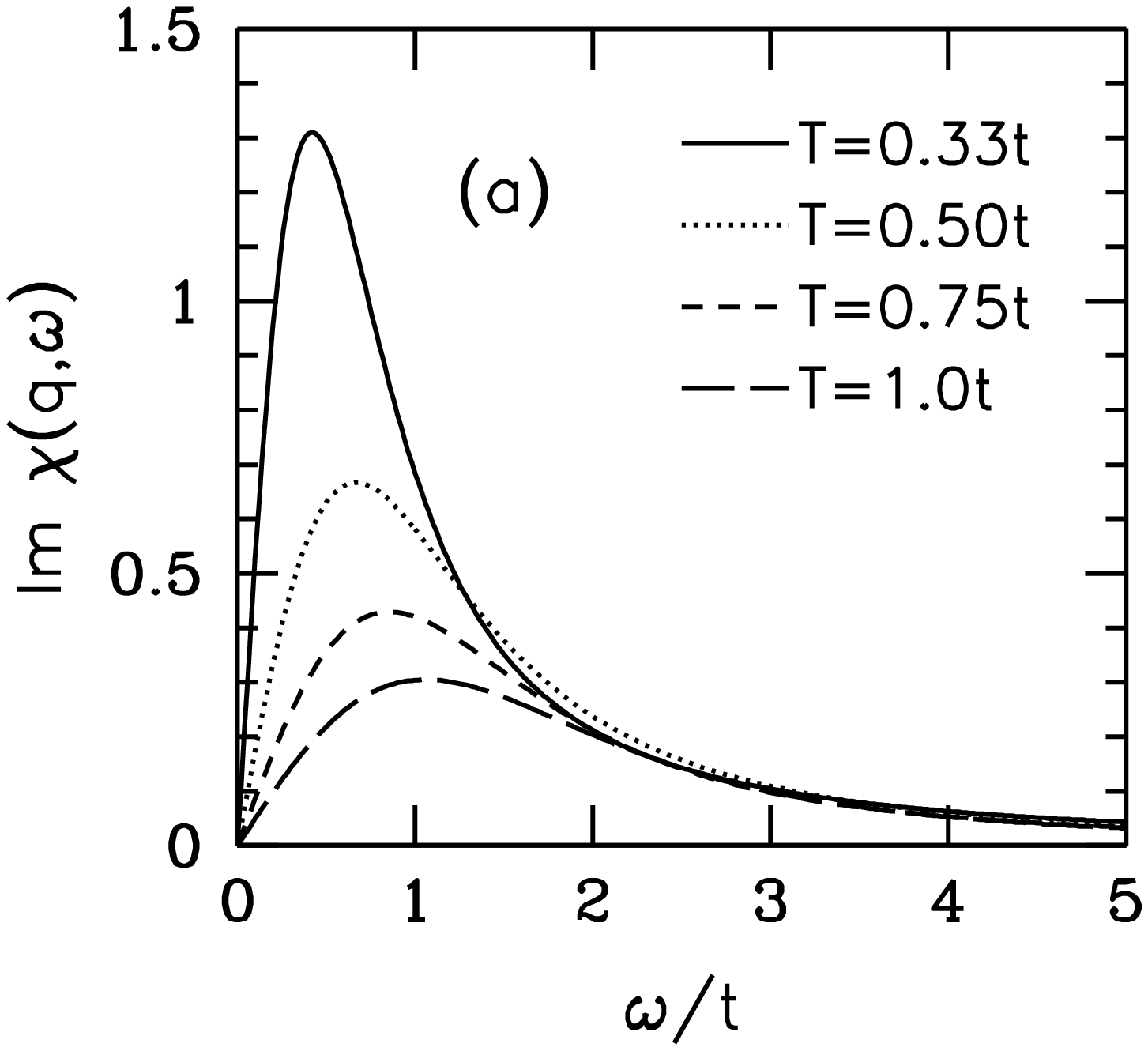}
\epsfysize=6.8cm \epsffile[98 184 672 598] {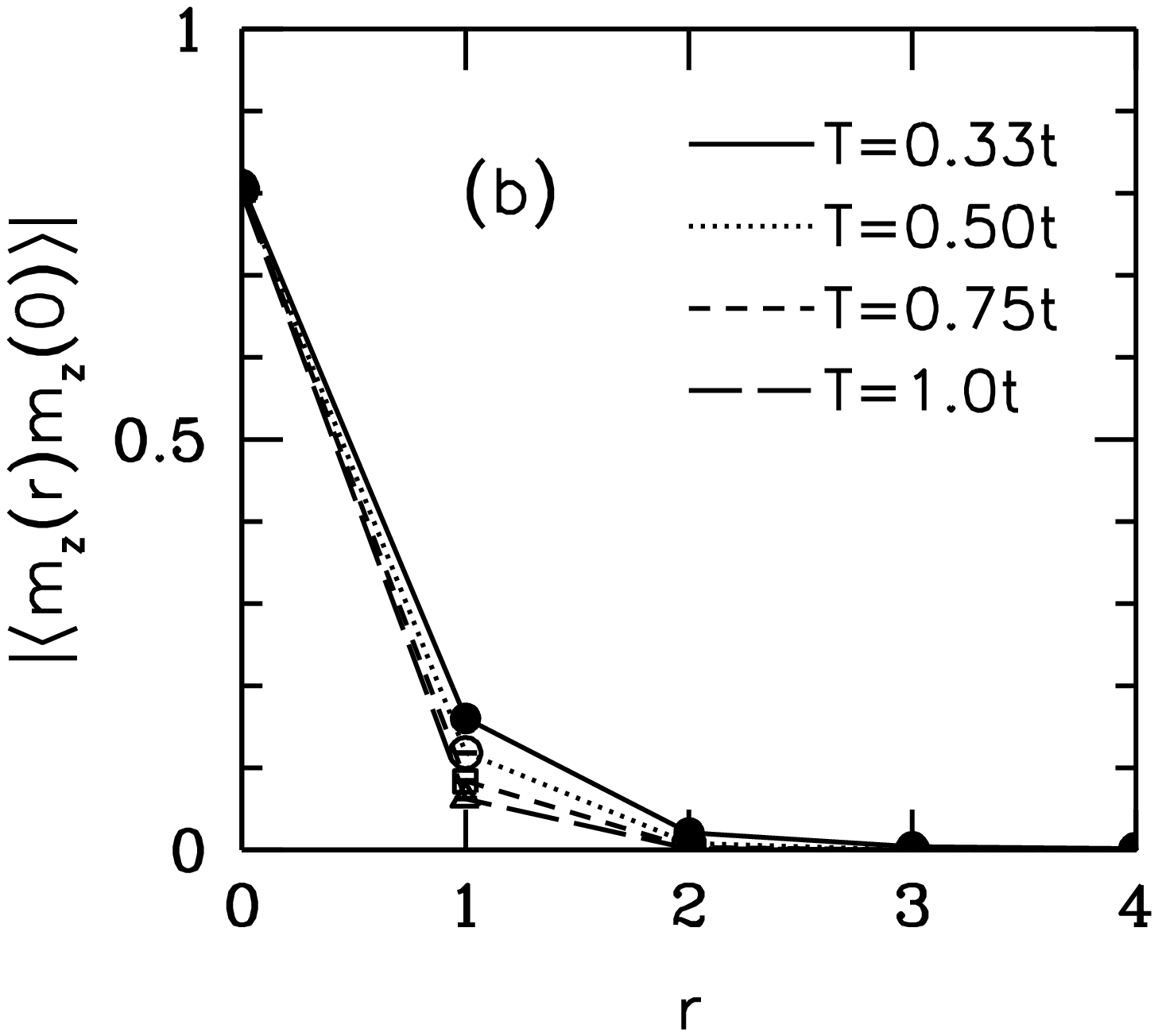}}
\vspace{0.3cm}
\caption{
(a) Frequency dependence of ${\rm Im}\,\chi(\q,\om)$ at
$\q=(\pi,\pi)$.
(b) $|\langle m_z(\r)m_z(0)\rangle |$
versus $\r$ along the $(1,0)$ direction.
These results are for $\xn=0.875$ and $U/t=8$.
\label{fig:imchi}}
\end{figure}

Using Quantum Monte Carlo simulations \cite{SRW}
we have measured the staggered
magnetic susceptibility
\begin{equation}
\chi({\bf q},i\omega_m) = {1\over N} \sum_{\lb}
  \int^\beta_0 d\tau\, 
   e^{i\omega_m\tau} 
   e^{-i\q\cdot\lb} 
     \left\langle m^-_{i+\ell}(\tau) m^+_{i}(0) \right\rangle.
\label{eq:chiqw}
\end{equation}
Here 
$m^+_i(0) = c^{\dagger}_{i\uparrow}c_{i\downarrow}$
and 
$m^-_{i+\ell}(\tau) = e^{H\tau} m^-_{i+\ell}(0) e^{-H\tau}$,
where $m^-_{i+\ell}(0)$ is the hermitian conjugate of 
$m^+_{i+\ell}(0)$.

It is useful to have estimates of the 
characteristic length and energy scales of the AF 
correlations.
For this purpose, in Fig. \ref{fig:chiq}(a) we first
show $\chi(\q,i\om_m=0)$ versus $\q$ for $\xn=0.875$
and $U/t=8$.
We see that as $T$ is lowered below $J\simeq 4t^2/U$,
strong AF correlations develop.
Next, in
Fig. \ref{fig:chiq}(b)
\begin{equation}
F(\q) =  
\lim_{\om\rightarrow 0}
{ {\rm Im}\,\chi(\q,\om) \over \omega}
\end{equation}
is shown.
Here, ${\rm Im}\,\chi(\q,\om)$ has been obtained by 
numerical analytic continuation \cite{ME} of $\chi(\q,i\om_m)$.
It is well known that 
$F(\q)$ is the quantity which determines the NMR $T_1^{-1}$ 
response of the system. 
We see that $F(\q\sim (\pi,\pi))$ increases rapidly
as $T$ is lowered.
The low frequency nature of the AF correlations are also 
seen in Fig. \ref{fig:imchi}(a), 
where ${\rm Im}\,\chi(\q=(\pi,\pi),\om)$ 
versus $\om$ is plotted.
In the cuprates, the longitudinal relaxation rate $T_1^{-1}$
of the planar Cu nuclei is dominated by the low frequency 
spin fluctuations near $\q=(\pi,\pi)$.
On the other hand, because of the form factor of the oxygen
hyperfine coupling,
the $T_1^{-1}$ of the oxygen nuclei is
mostly determined by the background of $F(\q)$
away from $(\pi,\pi)$.

In order to have an estimate of the 
characteristic length scales, in Fig. 2(b) we have plotted
the magnitude of the magnetization correlation function
$|\langle m_z(\r)m_z(0)\rangle |$ as a function of $\r$ 
along the $(1,0)$ direction.
We see that $|\langle m_z(\r)m_z(0)\rangle |$
decays very rapidly, and the range of the AF correlations are of 
order a lattice spacing at these temperatures.

\begin{figure} 
\centerline{\epsfysize=6.8cm \epsffile[-30 184 544 598] {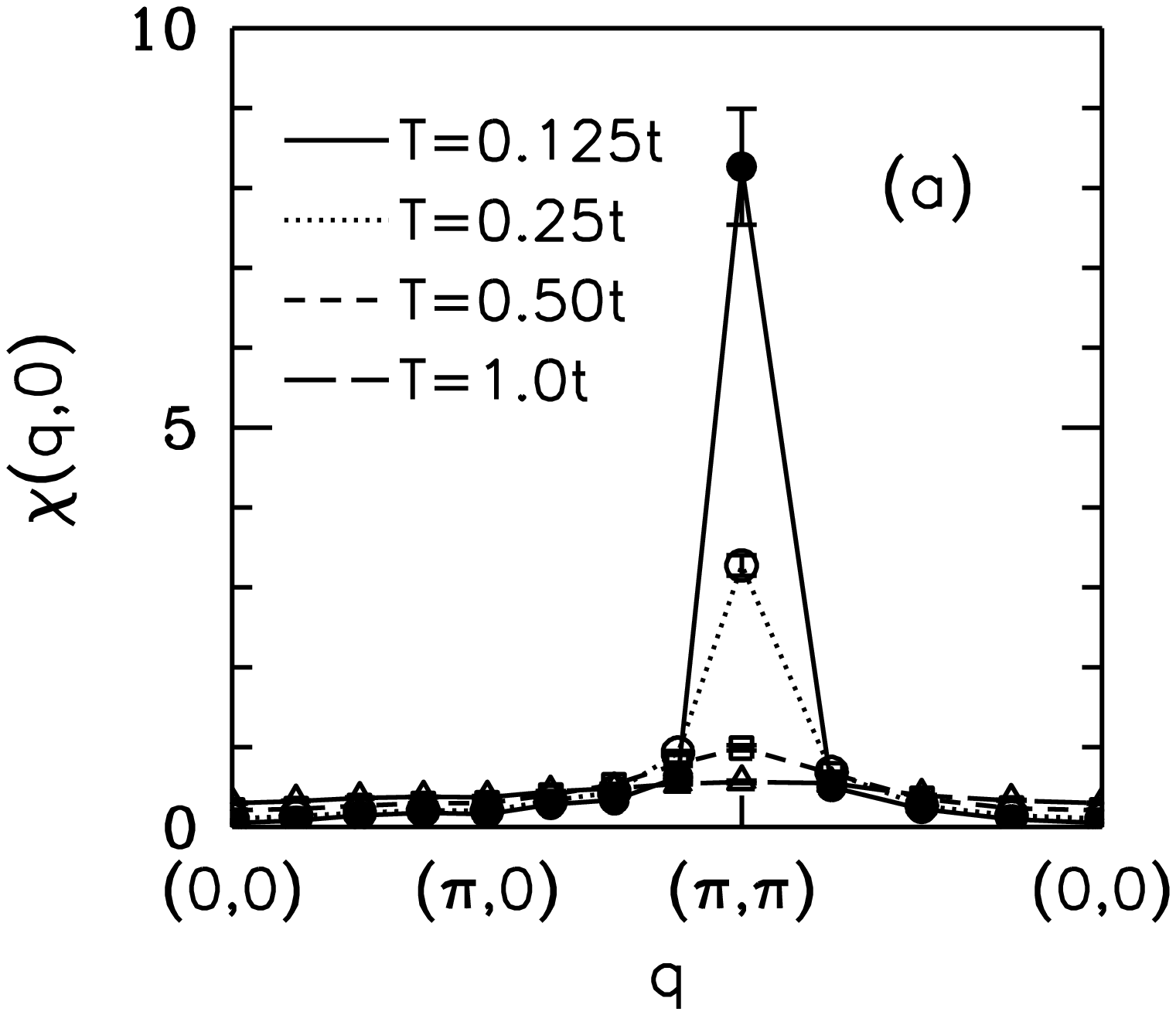}
\epsfysize=6.8cm \epsffile[98 184 672 598] {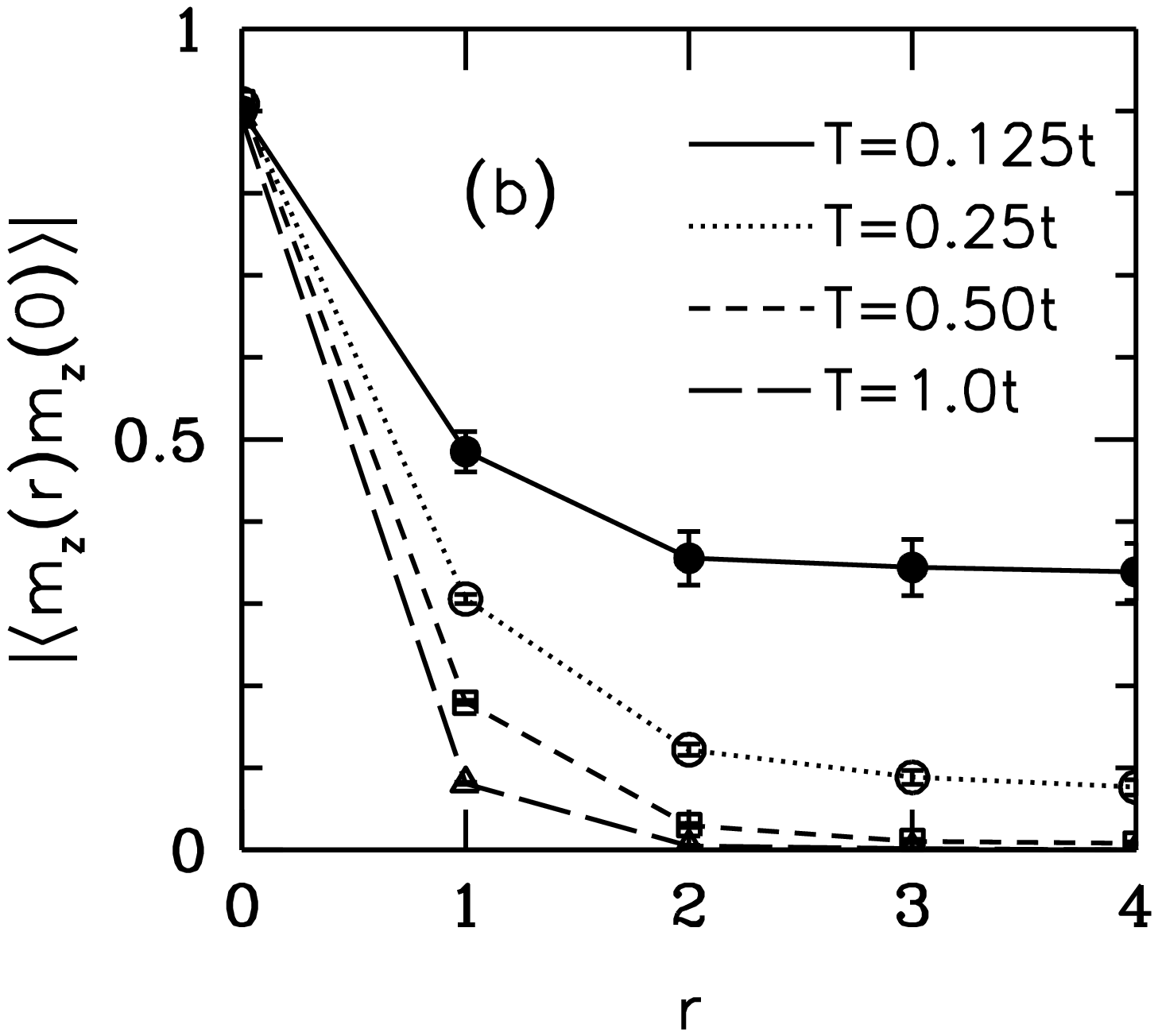}}
\vspace{0.3cm}
\caption{
(a) Momentum dependence of $\chi(\q,0)$.
(b) $|\langle m_z(\r)m_z(0)\rangle |$ versus $\r$.
These results are for $\xn=1.0$ and $U/t=8$.
\label{fig:chih}}
\end{figure}

These results demonstrate the existence of short--range
and low--frequency AF correlations in the metallic state near 
half--filling.
We note that while the AF correlations are short range, there is
significant spectral weight under the peak at 
$\q=(\pi,\pi)$ as seen in Fig. \ref{fig:chiq}.
It is interesting to compare these AF fluctuations with those 
at half--filling, where 
the ground state has long--range AF order \cite{Hirsch}.
Fig. \ref{fig:chih} shows results on $\chi(\q,0)$ and  
$|\langle m_z(\r)m_z(0)\rangle |$
at half--filling.
At $T=0.125t$, the long--range order is clearly 
established on the $8\times 8$ lattice.

\section{Charge Fluctuations}

\begin{figure} 
\centerline{\epsfysize=6.8cm \epsffile[-30 184 544 598] {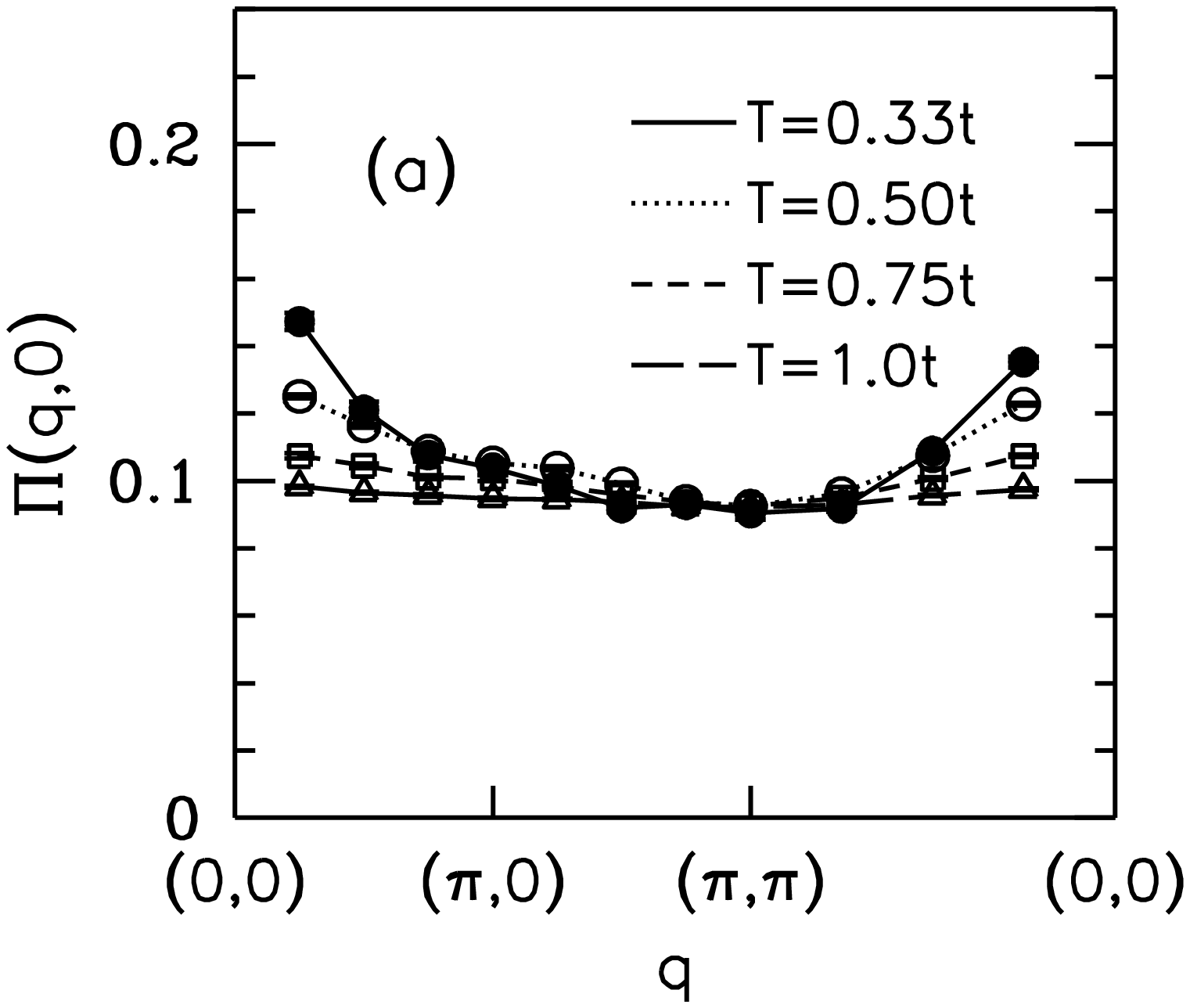}
\epsfysize=6.8cm \epsffile[98 184 672 598] {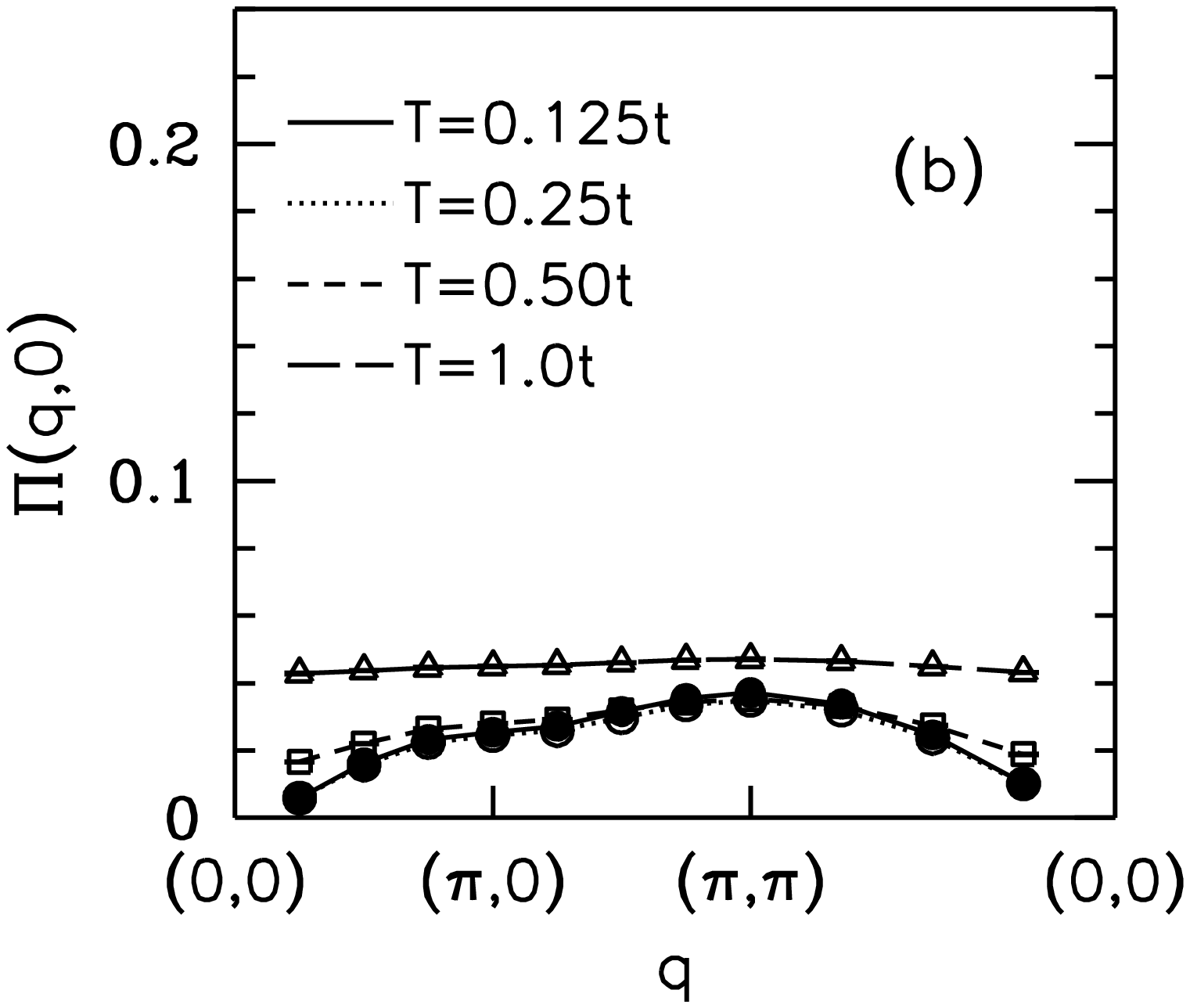}}
\vspace{0.3cm}
\caption{
Momentum dependence of $\Pi(\q,0)$ for $U/t=8$
at (a) $\xn=0.875$ and (b) 1.0.
\label{fig:pi}}
\end{figure}

\begin{figure}
\centerline{\epsfysize=6.8cm \epsffile[-30 184 544 598] {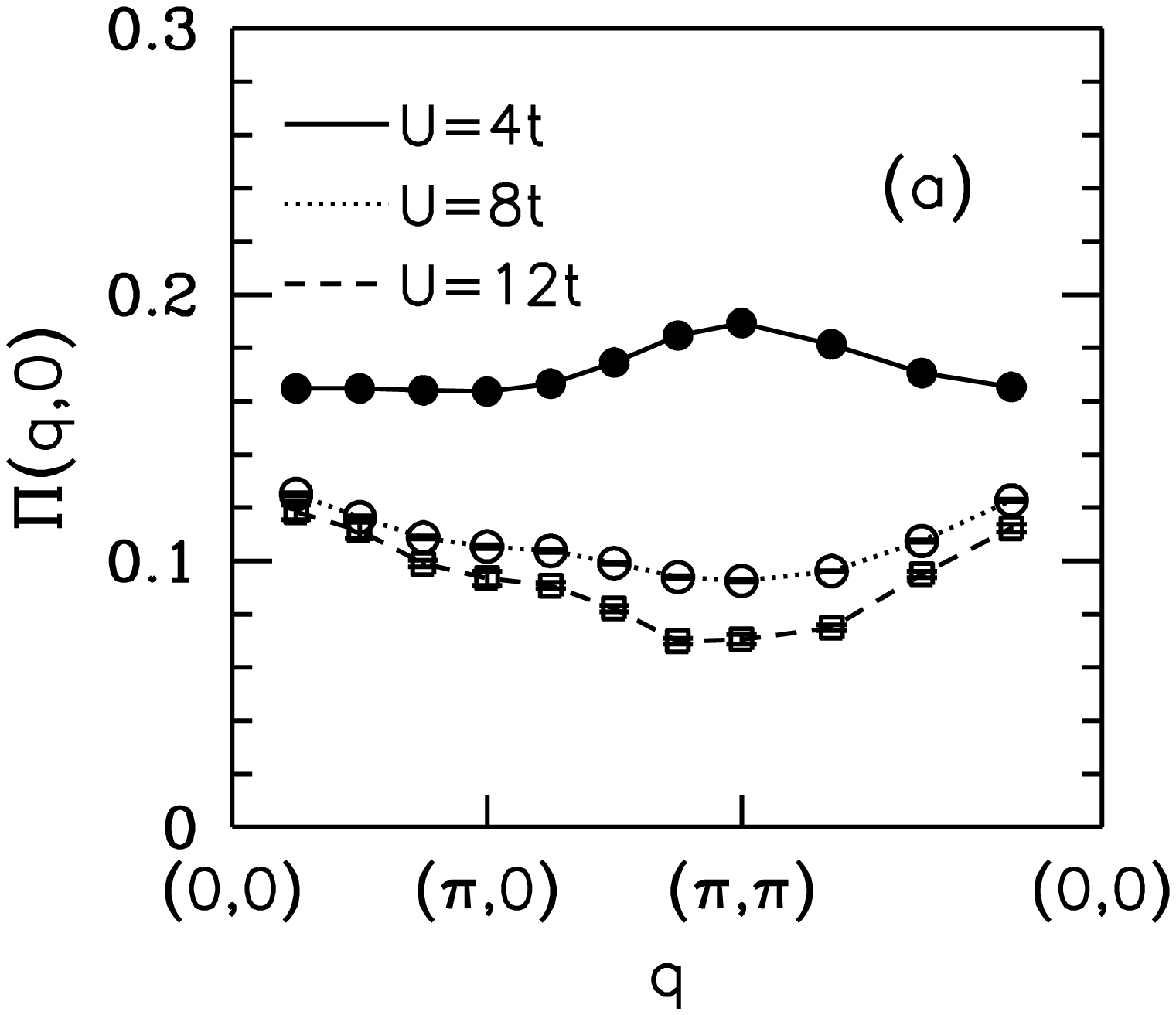}
\epsfysize=6.8cm \epsffile[98 184 672 598] {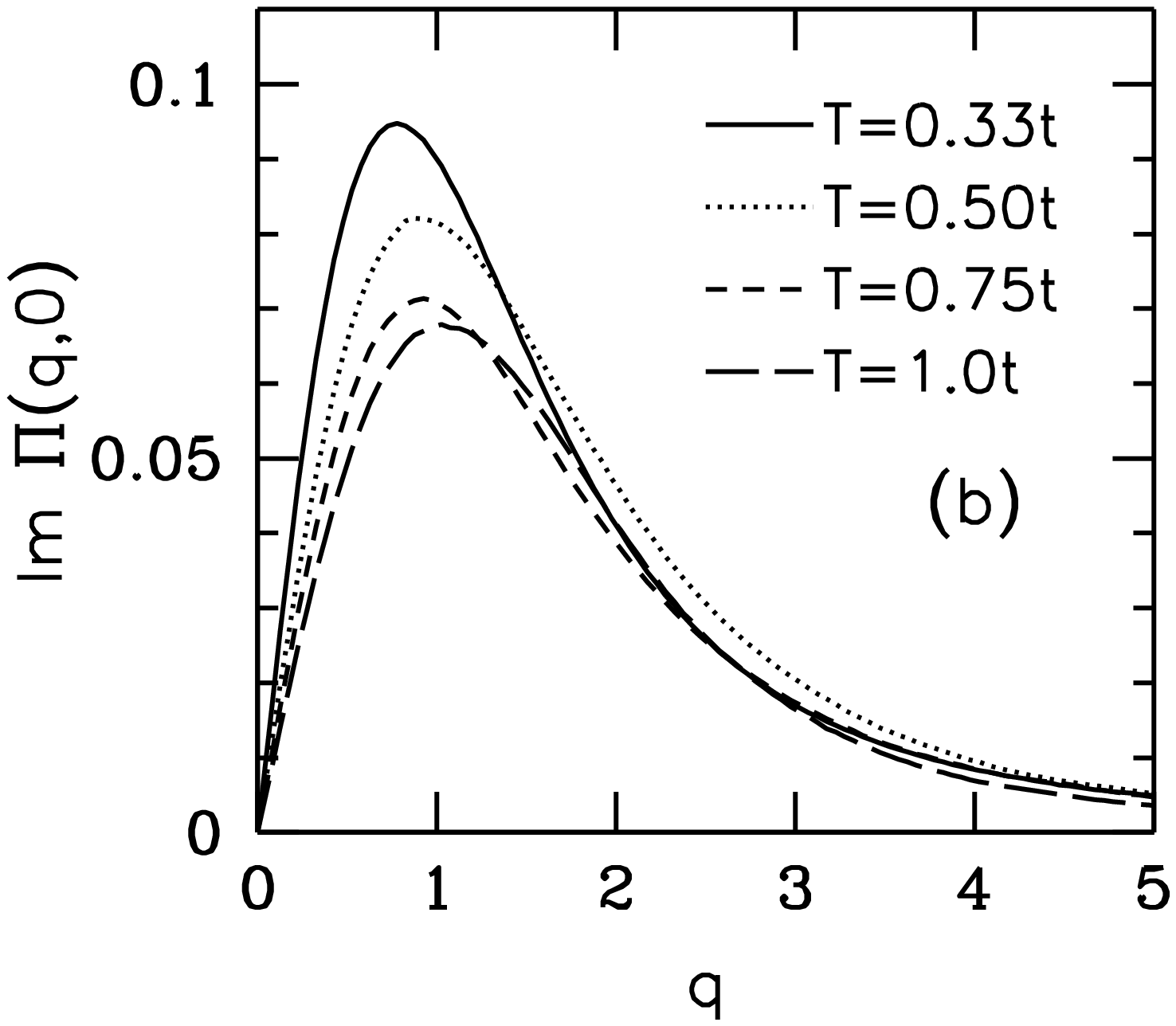}}
\vspace{0.3cm}
\caption{
(a) $\Pi(\q,0)$ versus $\q$ for various values of $U/t$
at $T/t=0.5$.
(b) ${\rm Im}\,\Pi(\q,\om)$ versus $\om$ at $\q=(\pi/4,0)$
and various temperatures for $U/t=8$.
These results are for $\xn=0.875$.
\label{fig:impi}}
\end{figure}

In this section, we study the dynamics of the charge fluctuations.
We will present results on the charge dynamics obtained from 
\begin{equation}
\Pi({\bf q},i\omega_m) = 
  \int^\beta_0 d\tau\, 
   e^{i\omega_m\tau} 
     \left\langle n(\q,\tau) n(-\q,0) \right\rangle,
\label{eq:piqw}
\end{equation}
where 
$n(\q,0)={1\over \sqrt{N}} 
\sum_{\p,\sigma} 
c^{\dagger}_{\p+\q\sigma} c_{\p\sigma}$.

Figures \ref{fig:pi}(a) and (b) show
$\Pi(\q,i\om_m=0)$ versus $\q$ for $\xn=0.875$ and 1.0.
The dependence of $\Pi(\q,0)$ on $U/t$ is shown in
Fig. \ref{fig:impi}(a) for $\xn=0.875$.
Fig. \ref{fig:impi}(b) shows 
the temperature evolution of the 
charge fluctuation spectral weight ${\rm Im}\,\Pi(\q,\om)$ 
for $\q=(\pi/4,0)$ and $\xn=0.875$.
In these figures,
we see that, on the average,
the onsite Coulomb repulsion suppresses the charge fluctuations.
At half--filling, $\Pi(\q\rightarrow 0,0)$ vanishes 
as the Mott--Hubbard gap opens.
However, for $\xn=0.875$, 
the $\q\rightarrow 0$ part actually gets enhanced,
as $T$ is lowered.

The $\q\rightarrow 0$ limit of $\Pi(\q,0)$ gives the 
compressibility $\kappa$ of the system.
The low--temperature Monte Carlo studies \cite{Imada}
find that the enhancement of the long wavelength
charge response near half--filling
is due to the proximity of the system 
to a metal--insulator transition,
and that, 
as the doping $\delta$ is reduced, $\kappa$ diverges as
$\delta^{-1}$.
The enhancement of the long wavelength charge response
has many experimental consequences.
For instance, the long wavelength phonon modes 
of a real material can get softened 
by the enhanced $\Pi(\q,\om)$.


\section{Single--Particle Excitations}

\begin{figure}
\centerline{\epsfysize=8cm \epsffile[18 184 592 598] {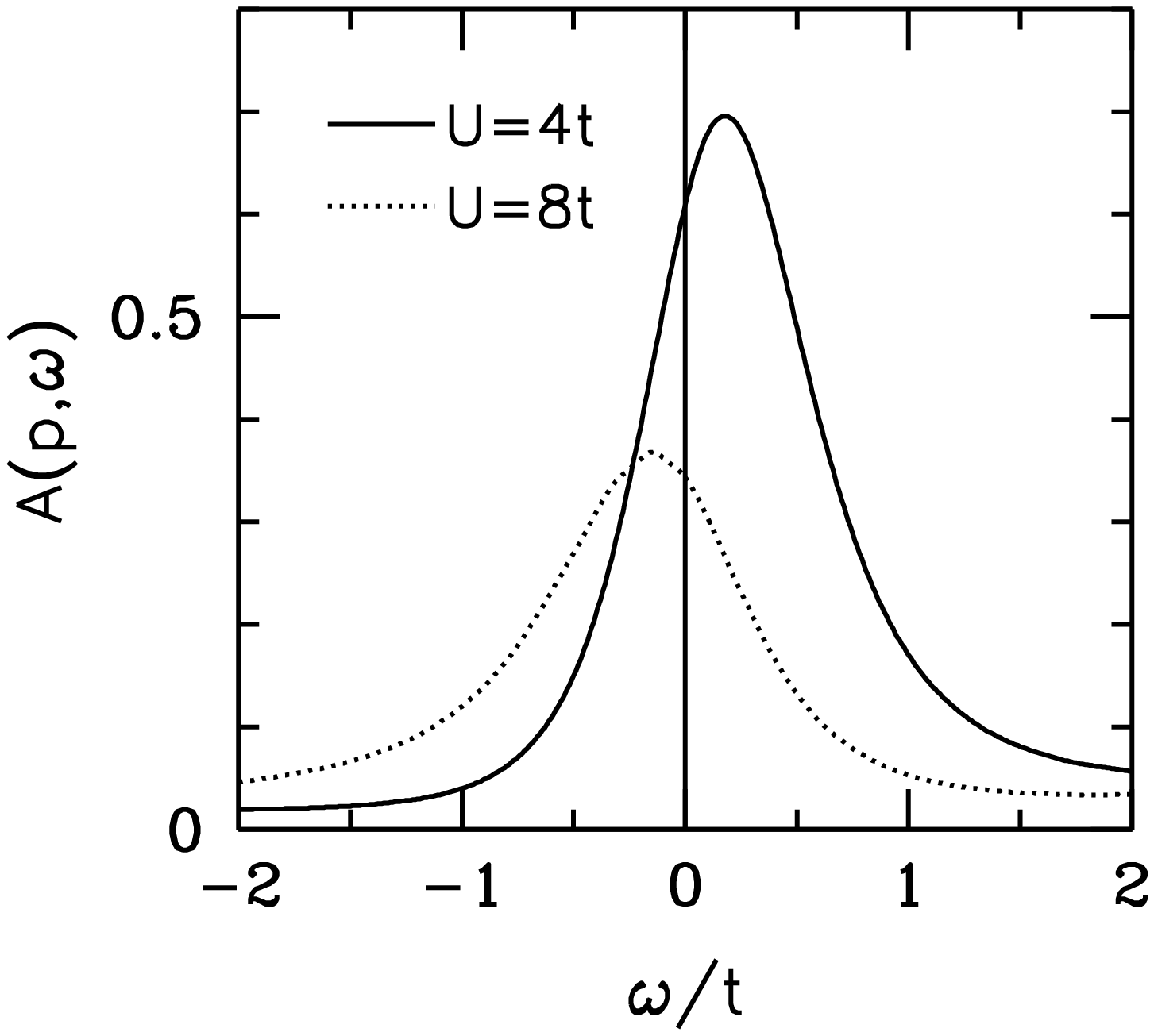}}
\vspace{0.3cm}
\caption{
Single--particle spectral weight $A(\p,\om)$ versus $\om$
for $U/t=4$ and 8 at $\xn=0.875$ and $T=0.33t$.
\label{fig:apw}}
\end{figure}

\begin{figure} 
\centerline{\epsfysize=6.8cm \epsffile[-30 184 544 598] {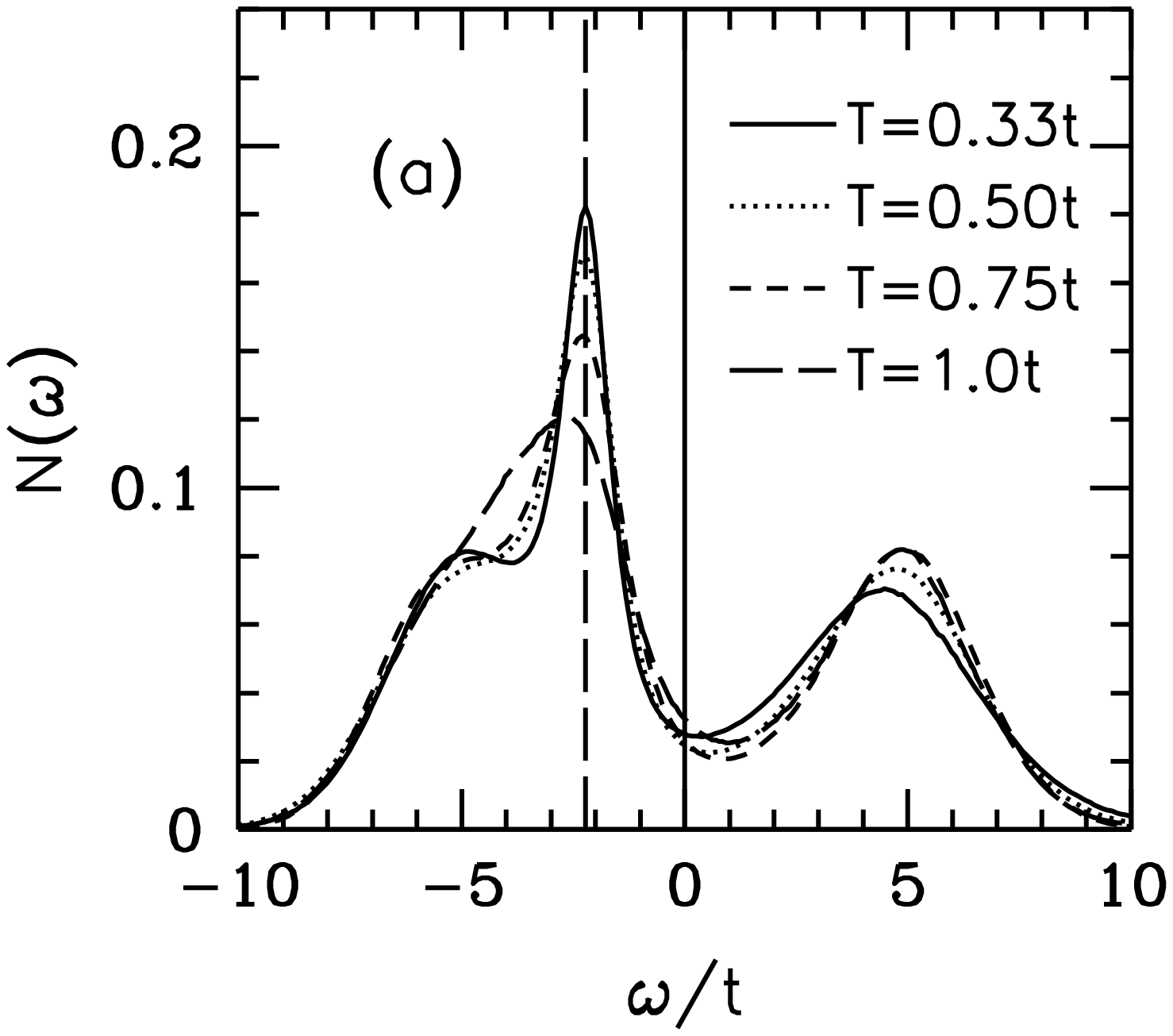}
\epsfysize=6.8cm \epsffile[98 184 672 598] {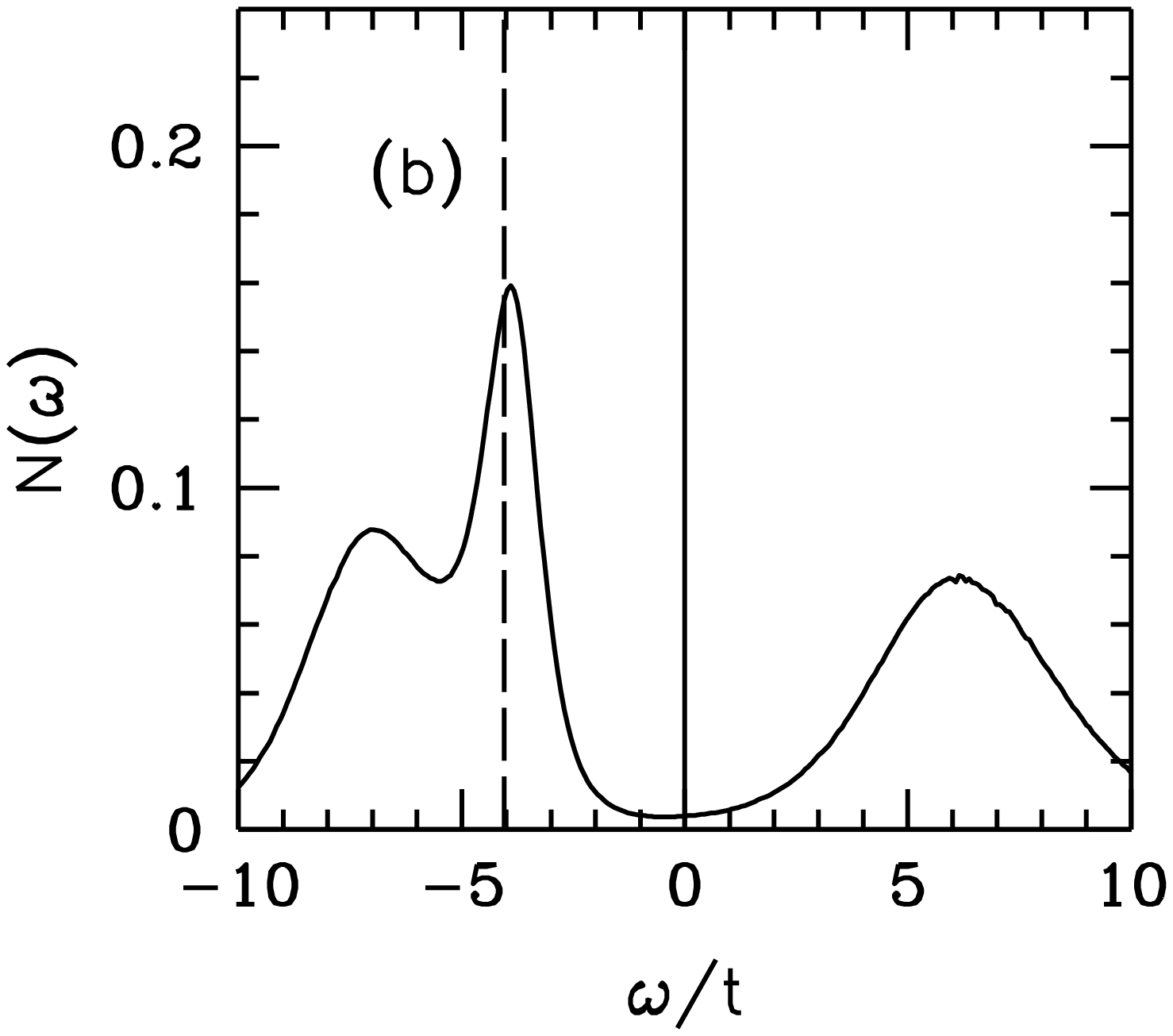}}
\vspace{0.3cm}
\caption{
(a) Temperature evolution of $N(\om)$ versus $\omega/t$ for $U/t=8$.
(b) $N(\om)$ versus $\om$ for $U/t=12$ and $T=0.5t$.
These results are for $\xn=0.875$.
\label{fig:Nw}}
\end{figure}

\begin{figure}
\centerline{\epsfysize=8cm \epsffile[18 184 592 598] {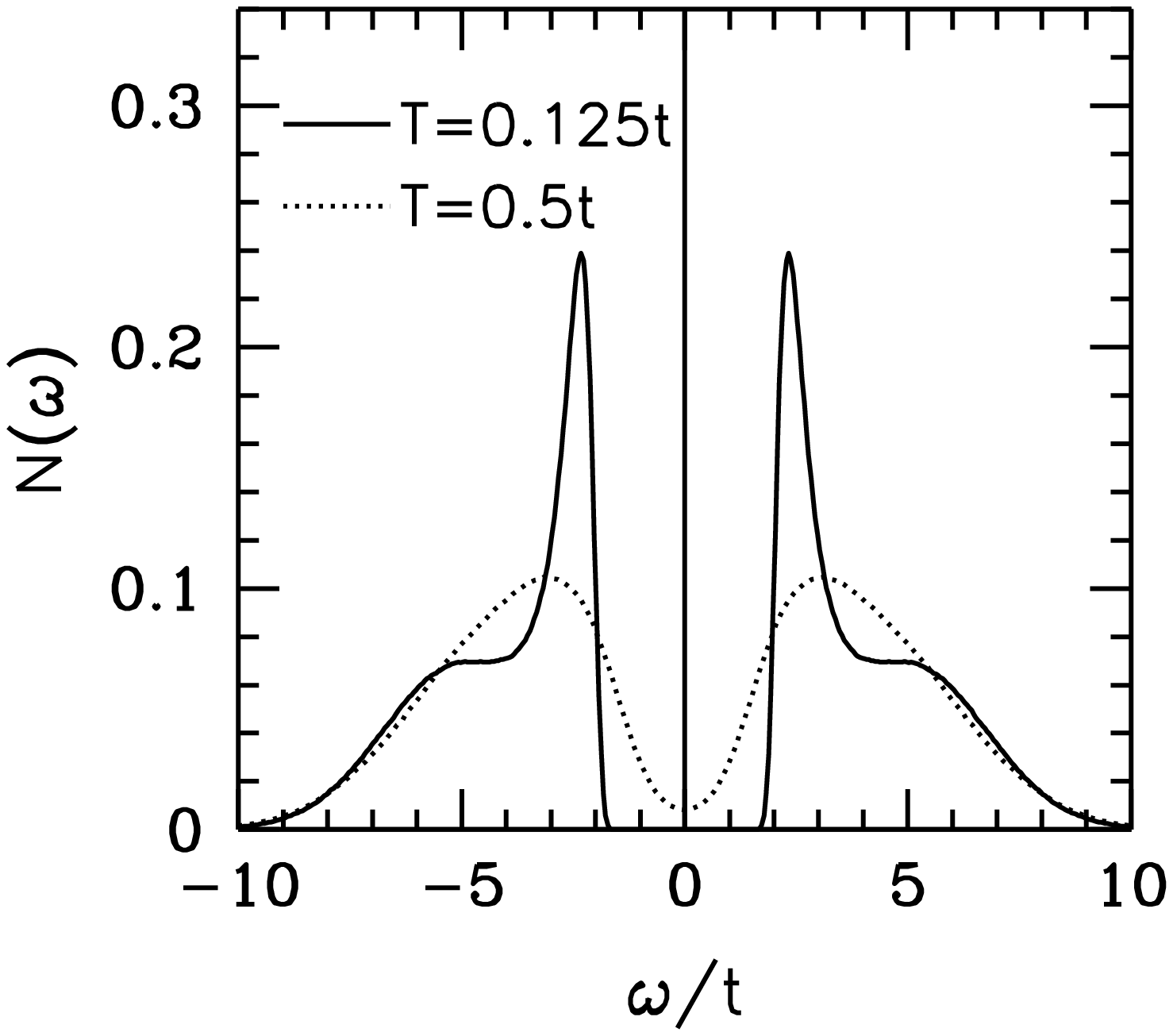}}
\vspace{0.3cm}
\caption{
$N(\om)$ versus $\om$ at half--filling for $U/t=8$.
\label{fig:Nwh}}
\end{figure}

In this section, we review numerical results on the single--particle 
excitations \cite{Lancapw,Bul3,Moreo}.  
We are especially interested in the effects of the 
spin and charge fluctuations 
on the single--particle excitations.
The single--particle spectral weight and the 
density of states are given by 
\begin{equation}
A({\bf p},\omega) = -{1\over \pi}\,\im\,G({\bf p},
i\omega_n\to\omega+i\delta) 
\label{eq:apw}
\end{equation}
and 
\begin{equation}
N(\omega) = {1\over N}
\sum_{\p} A({\bf p},\omega).
\label{eq:nw}
\end{equation}

In order to illustrate how the Coulomb repulsion affects the 
single--particle properties, 
in Fig. \ref{fig:apw} we show 
$A(\p=(\pi,0),\om)$ versus $\om$ for 
$\xn=0.875$, $T=0.33t$, and $U/t=4$ and 8.
Next, the temperature evolution of $N(\om)$ is shown in
Fig. \ref{fig:Nw}(a).
The main features of the spectrum are lower and upper Hubbard
bands, and a narrow metallic quasiparticle band at the top of the
lower Hubbard band.
Surprisingly, 
the quasiparticle band has strong $T$ dependence.
Comparing this figure with 
Figures \ref{fig:chiq} and \ref{fig:pi}(a),
we see that the temperature evolution of the quasiparticle band
coincides with the development 
of the $\q\sim(\pi,\pi)$ magnetic and $\q\sim (0,0)$ charge fluctuations.
Fig. \ref{fig:Nw}(b) shows $N(\om)$ for $U/t=12$ and
$T/t=0.5$, where the upper Hubbard band is further split from 
the lower Hubbard and quasiparticle bands.

Results for $N(\om)$ at half--filling are shown in 
Fig. \ref{fig:Nwh}.
At $T=0.5t$, we observe a Mott--Hubbard pseudogap in the spectrum.
At $T=0.125t$, where long--range order is established on the 
$8\times 8$ lattice, a full single--particle gap is seen.
At the moment, it is not completely understood
how in the ground state of the sytem
the metallic band, which is seen developing in 
Fig. \ref{fig:Nw}(a), evolves to the 
$N(\om)$ of the insulating state as the doping is reduced.
At the temperatures that the Monte Carlo calculations are carried out, 
a metallic band forms at the top of the lower Hubbard band even for 
3\% doping.

\begin{figure}
\centerline{\epsfysize=8cm \epsffile[18 184 592 598] {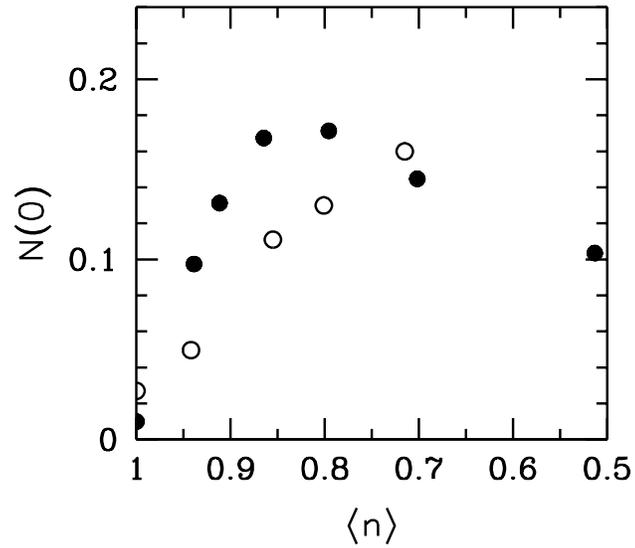}}
\vspace{0.3cm}
\caption{
Density of states at the fermi level $N(0)$ versus the filling
$\xn$ for $U/t=8$, and
$T/t=0.5$ (solid circles) and 1.0 (open circles).
\label{fig:Nfill}}
\end{figure}

Figure \ref{fig:Nfill} shows the filling dependence of 
the density of states at the fermi level
$N(0)$ for $U/t=8$.
At $T=0.5t$, the filling where the maximum of 
$N(0)$ occurs is near $\xn=0.85$.
However, the temperature evolution of the Monte Carlo data 
indicates that this critical filling will move closer to half--filling
as $T\rightarrow 0$.
In fact, for the simplest Hubbard model that we are using 
where only the near--neighbor hopping $t$ and the onsite 
Coulomb repulsion $U$ are taken into account,
it might occur infinitesimally close to half--filling.
Furthermore,
simple additional terms in the Hubbard Hamiltonian,
such as a next--near--neighbor hopping $t'$, might also
influence the position of the maximum of $N(0)$.


\section{Pairing Correlations}

\begin{figure} 
\begin{picture}(30,15)
\end{picture}
\includegraphics{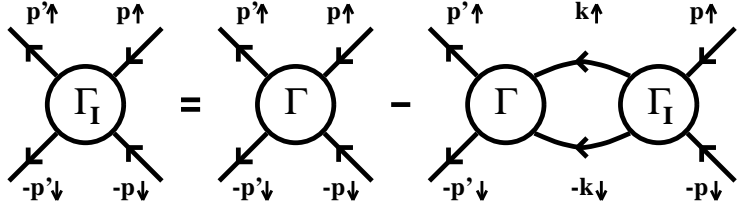}
\vspace{3.5cm}
\caption{
$t$--matrix equation relating the reducible and the irreducible 
interactions in the BCS channel.
\label{fig:t}}
\end{figure}

In this section, 
we will present results on the effective particle--particle 
interaction and the solution of the Bethe--Salpeter
equation in the BCS channel.

\begin{figure}
\centerline{\epsfysize=6.8cm \epsffile[-30 184 544 598] {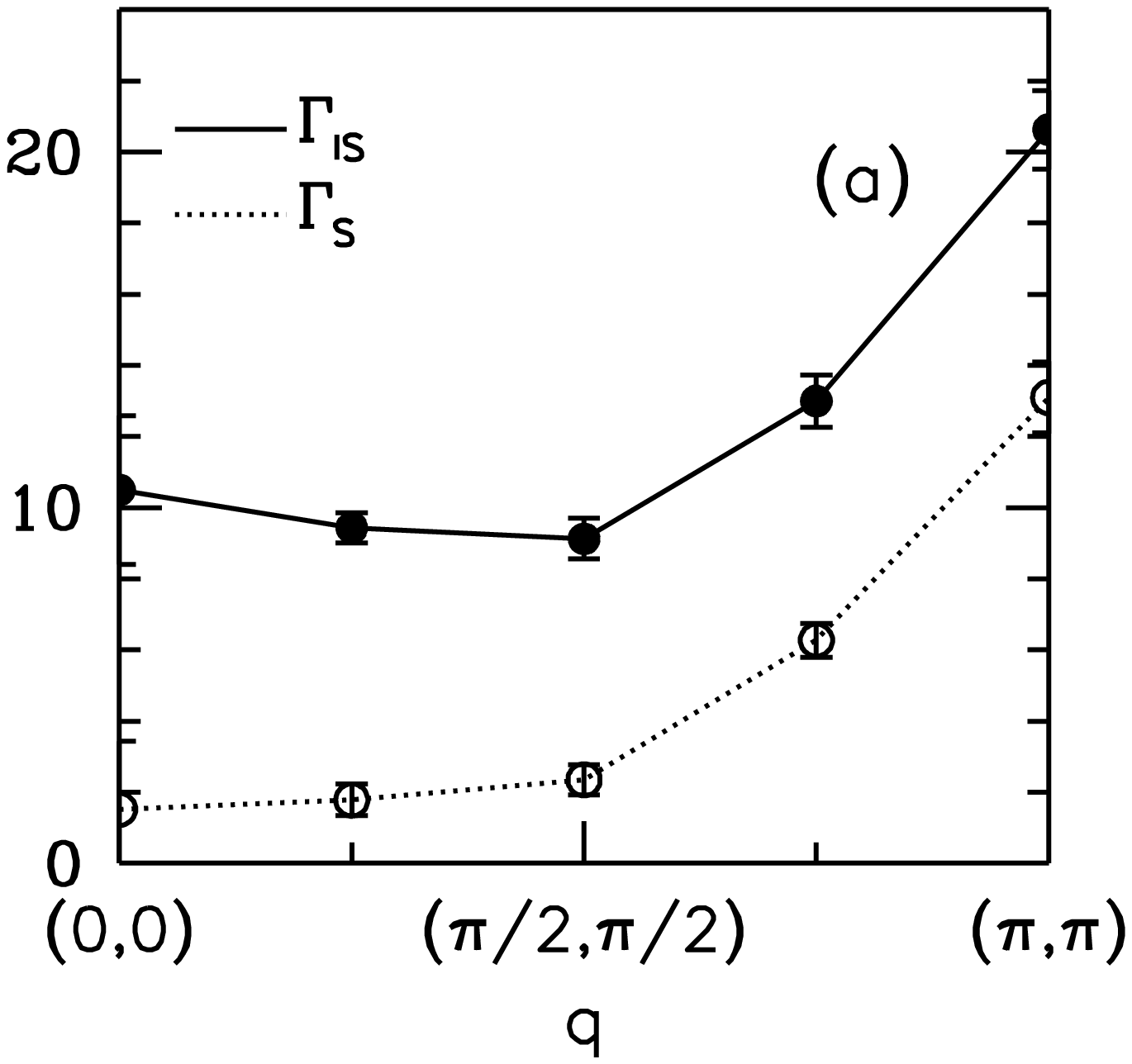}
\epsfysize=6.8cm \epsffile[98 184 672 598] {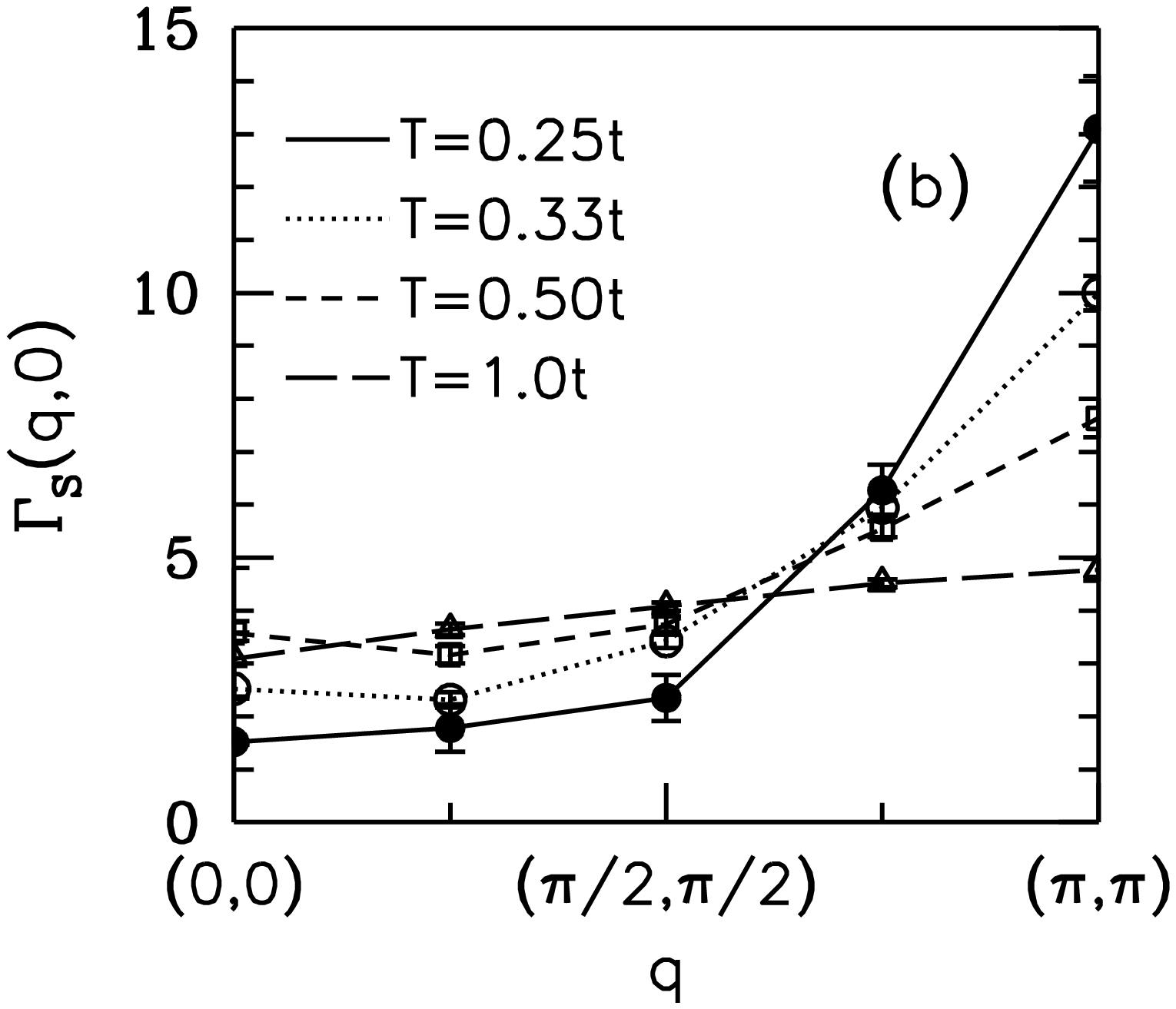}}
\vspace{0.3cm}
\caption{
(a) Momentum dependence of the 
particle--particle reducible and irreducible interactions 
in the singlet channel,
$\Gamma_{\rm S}$ and $\Gamma_{\rm IS}$,
at $T/t=0.25$.
(b) Temperature evolution of $\Gamma_{\rm S}(\q,0)$ versus $\q$.
These results are for $U/t=4$ and $\xn=0.875$.
\label{fig:Gamma4}}
\end{figure}

Using Quantum Monte Carlo simulations we have calculated the reducible 
particle--particle interaction $\Gamma(p',-p',p,-p)$, which we will denote
by $\Gamma(p'|p)$ \cite{Effective}.
Here, $p$ stands for both $\p$ and $i\om_n$.
The reducible interaction is related to the irreducible interaction
$\Gamma_{\rm I}$ through the $t$--matrix equation shown
in Fig. \ref{fig:t}.
In the following, we will present results in the singlet channel where
$\Gamma_{\rm S}(p'|p)={1\over 2} (\Gamma(p'|p)+\Gamma(-p'|p))$.
Figure \ref{fig:Gamma4}(a) shows the momentum dependence of 
$\Gamma_{\rm S}(p'|p)$ and $\Gamma_{\rm IS}(p'|p)$
at $\om_n=\om_{n'}=\pi T$, corresponding to an energy transfer 
$\om_m=0$, for $U/t=4$ and $\xn=0.875$.
In this figure $\p$ is fixed at $(\pi,0)$ and 
$\q=\p'-\p$ is swept along the $(1,1)$ direction.
We observe that both $\Gamma_{\rm S}$ and $\Gamma_{\rm IS}$ peak
at large momentum transfers near $\q=(\pi,\pi)$.
The difference between 
$\Gamma_{\rm S}$  and $\Gamma_{\rm IS}$ 
represents the effects of the repeated particle--particle scatterings
in the BCS channel.
Figure \ref{fig:Gamma4}(b) shows $\Gamma_{\rm S}(\q,0)$ versus
$\q$ at various temperatures.
We see that, as $T$ is lowered,
the $\q\sim(\pi,\pi)$ part of $\Gamma_{\rm S}$ gets enhanced,
while the $\q\sim(0,0)$ part decreases.

\begin{figure} 
\centerline{\epsfysize=6.8cm \epsffile[-30 184 544 598] {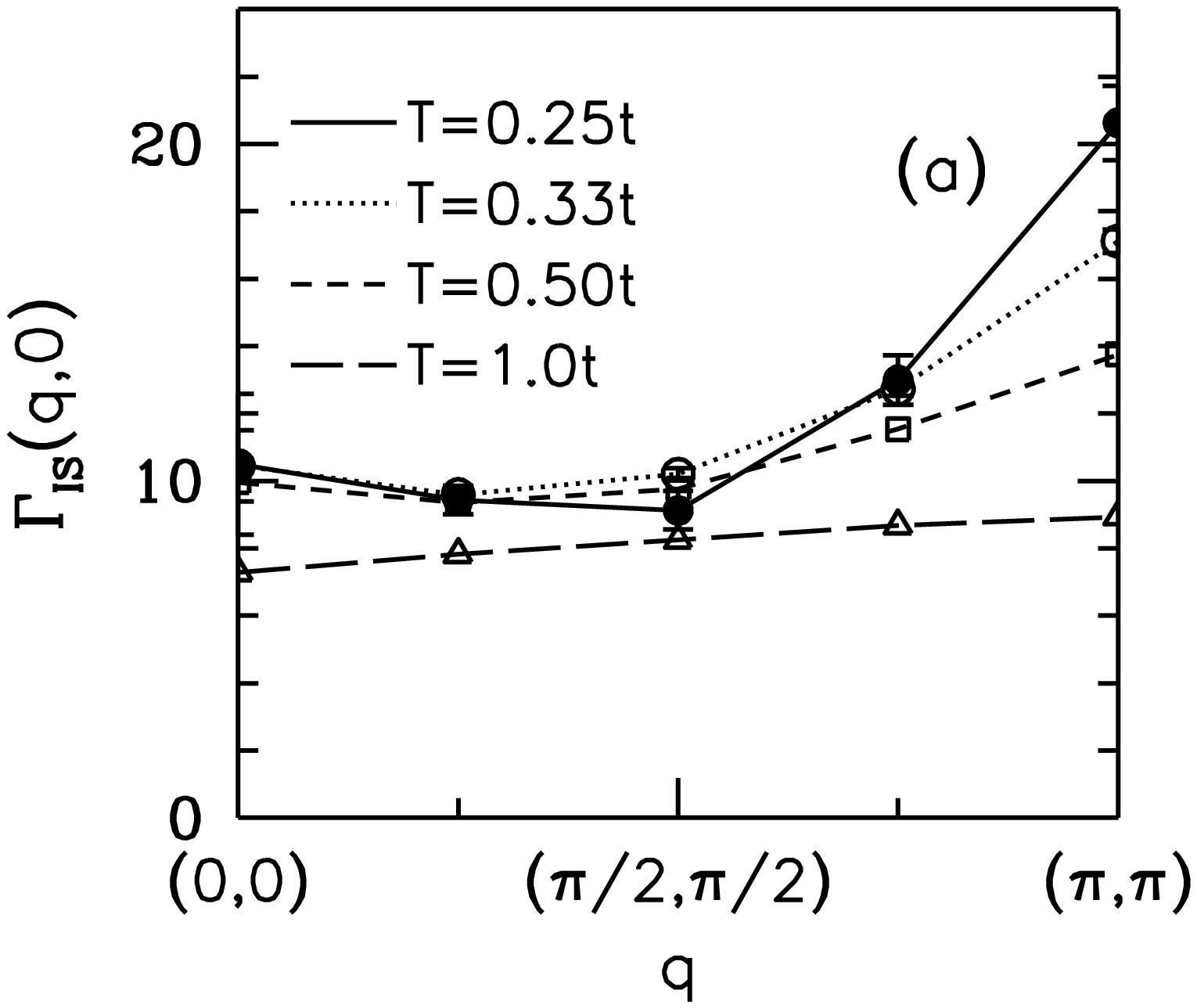}
\epsfysize=6.8cm \epsffile[98 184 672 598] {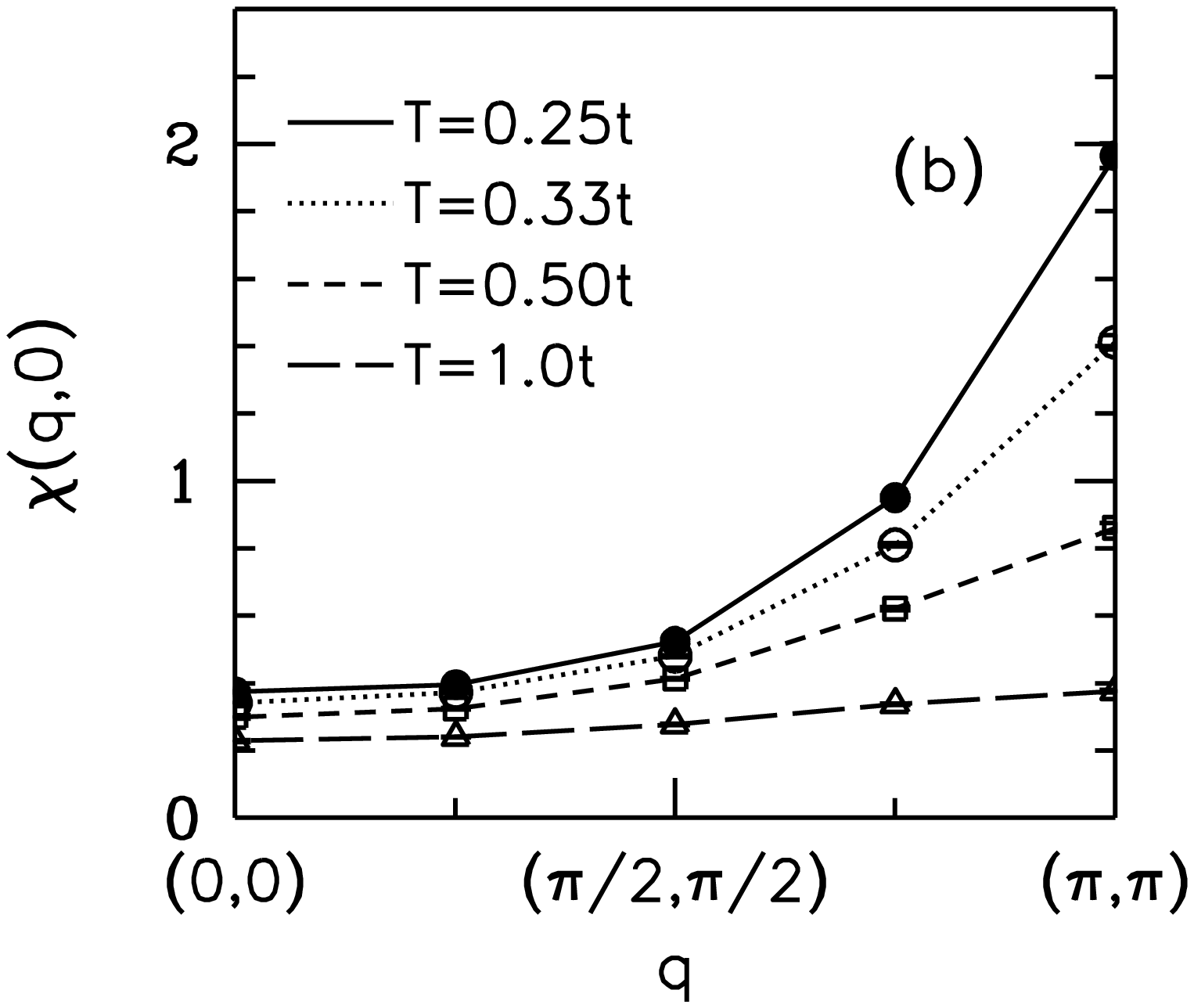}}
\vspace{0.3cm}
\caption{
Momentum dependence of (a) $\Gamma_{\rm IS}(\q,0)$ and 
(b) $\chi(\q,0)$ at various temperatures
for $U/t=4$ and $\xn=0.875$.
\label{fig:GammaI}}
\end{figure}

Figure \ref{fig:GammaI} compares the temperature evolution of 
$\Gamma_{\rm IS}(\q=\p'-\p,i\om_m=0)$ with that of 
$\chi(\q,0)$ for $U/t=4$.
We find that 
$\Gamma_{\rm IS}(\q,0)$ shown in Fig. \ref{fig:GammaI}(a)
can be fit exceptionally well with a 
phenomenological spin--fluctuation exchange form given by
\begin{equation}
\Gamma_{\rm I}(q) = U + {3\over 2} (gU)^2\chi(q),
\end{equation}
where the renormalization coefficient $g\simeq 0.8$.

\begin{figure}
\centerline{\epsfysize=8cm \epsffile[18 184 592 598] {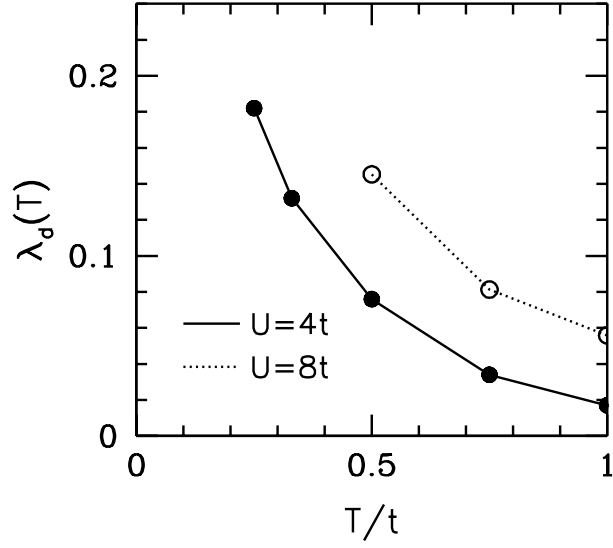}}
\vspace{0.3cm}
\caption{
Singlet $d_{x^2-y^2}$ eigenvalue versus $T/t$ 
at $\xn=0.875$.
\label{fig:lambda}}
\end{figure}

\begin{figure}
\centerline{\epsfysize=8cm \epsffile[18 184 592 598] {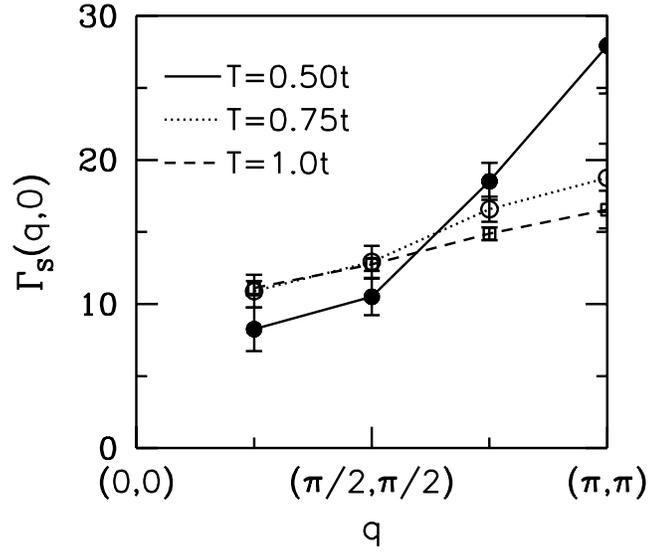}}
\vspace{0.3cm}
\caption{
Temperature evolution of the singlet reducible vertex 
$\Gamma_{\rm S}(\q,i\om_m=0)$ versus $\q$ for
$U/t=8$ and $\xn=0.875$.
\label{fig:Gamma8}}
\end{figure}

$\Gamma_{\rm IS}$ can be used, 
along with the single--particle Green's function $G(\p,i\om_n)$, 
to solve the Bethe--Salpeter equation 
\begin{equation}
\lambda_\alpha \phi_\alpha(p) =
-{T\over N} \sum_{p'} \Gamma_{\rm IS} (p|p') G_\uparrow(p')
G_\downarrow(-p') \phi_\alpha(p').
\label{eq:bs}
\end{equation}
The irreducible interaction $\Gamma_{\rm IS}$ is attractive for the 
solutions which have positive eigenvalues.
If the maximum eigenvalue reaches 1, then this signals 
a particle--particle instability of the system to a superconducting
state.
At the temperatures that the Monte Carlo calculations are 
carried out, the eigenvalues are much smaller than 1, and the 
system is far from a superconducting ground state, if one 
exists at all.
Nevertheless, one can still study which pairing channels are favored as 
the AF fluctuations develop.
We find that, in this regime, the singlet pairing correlations are 
strongest in the $d_{x^2-y^2}$ channel.
The solid circles in Fig. \ref{fig:lambda}
show the $d_{x^2-y^2}$ eigenvalue as a function of temperature for 
$U/t=4$ and $\xn=0.875$.

For $U/t=8$, we have not been able to calculate the irreducible 
interaction because of numerical convergence problems,
however we have calculated the reducible interaction and the 
Bethe--Salpeter eigenvalues.
Figure \ref{fig:Gamma8} shows the temperature evolution of the 
reducible interaction.
Just as for $U/t=4$, $\Gamma_{\rm S}(\q,i\om_m=0)$ 
at large momentum transfers is strongly repulsive,
and it grows as $T$ is lowered.
Consequently, 
the leading singlet pairing correlations occur in the 
$d_{x^2-y^2}$ channel.
The open circles in Fig. \ref{fig:lambda} represent 
the $d_{x^2-y^2}$ eigenvalue for $U/t=8$.

Finally, we would like to note that the $d_{x^2-y^2}$
pairing correlations do not necessarily require a long AF correlation
length, but simply large weight in $\Gamma_{\rm IS}$ 
at momentum transfers near $\q=(\pi,\pi)$.
For instance, 
we have seen in Fig. \ref{fig:imchi}(b) that 
the AF correlation length 
for $\xn=0.875$, $U/t=8$ and $T/t=0.33$
is only of order one lattice spacing.

\section{Conclusions}

Numerical studies of the Hubbard model find that a correlated
metallic band forms with unusual physical properties 
upon doping of the insulating half--filled system.
Here, we have seen that in this metallic state the 
single--particle excitations are 
strongly modified by the many--body effects,
the long--wavelength charge response is enhanced,
and the system has strong short--range AF fluctuations.
Also associated with this state are $d_{x^2-y^2}$ 
pairing correlations.

These electronic properties are similar to those of the
layered cuprates.
Angular resolved photoemission experiments 
on the cuprates
find that the quasiparticles are 
damped and renormalized.
NMR and inelastic neutron scattering experiments show
the existence of short--range AF correlations in the 
superconducting samples.
Furthermore,
a large number of experiments indicate
that the superconducting order parameter has 
$d_{x^2-y^2}$ symmetry.
For these reasons, we think that simple strongly correlated 
models such as the two--dimensional Hubbard model are useful in
understanding the unusual electronic properties of 
these materials.

\section*{Acknowledgments}

Tha author would like to thank D.J. Scalapino
for many helpful suggestions and discussions.
The work presented here was carried out in 
collaboration with D.J. Scalapino and S.R. White.
The author gratefully acknowledges support by the 
National Science Foundation under grant No.~DMR92--25027.
The numerical computations reported in this paper
were carried out at the San Diego Supercomputer Center.


\end{document}